\documentclass[letterpaper,12pt]{article}   
\usepackage{osajnl2} 
\usepackage[draft]{hyperref} 

\begin{document}

\title{Soliton Propagation with Cross Phase Modulation in Silicon Photonic Crystal Waveguides }

\author{Matthew Marko,$^{1,2,*}$ Xiujian Li,$^{2,3}$ and Jiangjun Zheng$^{2}$}
\address{$^1$Navy Air Warfare Center Aircraft Division (NAWCAD), Joint Base McGuire-Dix-Lakehurst, Lakehurst NJ 08733, USA}
\address{$^2$Department of Mechanical Engineering, Columbia University in the City of New York, New York NY 10027, USA}
\address{$^3$Tech-Physical Research Center Science College, National University of Defense Technology, Changsha, Hunan 410073, PRC}
\address{$^*$Corresponding author: matthew.marko@navy.mil}

\begin{abstract}
An effort was conducted to numerically determine, using the Nonlinear Schr\"{o}dinger Split-Step Fourier method, if using cross phase modulation could cause temporal soliton pulse propagation in a silicon slow-light photonic crystal waveguide shorter than a millimeter.  The simulations demonstrated that due to the higher powers and shorter scales of photonic crystals, two-photon absorption would cause an optical soliton pulse to be extremely dissipative.  The model demonstrated, however, that by utilizing cross-phase modulation, it is possible to sustain a compressed soliton pulse within a silicon photonic crystal waveguide subjected to two-photon absorption over longer relative distances.  
\end{abstract}


\maketitle

\section{Introduction}
Optical solitons are undistorted standing waves that result as part of the interplay between the nonlinear Kerr effect and anomalous dispersion \cite{1,2}.  One method of studying solitons numerically is to use the Split-Step Fourier Nonlinear Schr\"{o}dinger Equation (NLSE) \cite{1,2,3,4}.  The NLSE is one of the most effective methods for numerically analyzing soliton pulse propagation in a nonlinear dispersive waveguide.  

Photonic crystal waveguides (PhCWG) have extraordinary nonlinear effects \cite{6}, where measurements of the dispersion have shown an increase of five orders of magnitude in the Group Velocity Dispersion (GVD) compared to the material dispersion \cite{7,8}.  PhCWG, with their vastly increased nonlinear effects, are an ideal medium for reducing the nonlinear length scales from meters and kilometers typical of fibers, to sub-millimeter PhCWG.  A soliton is formed by the interplay of anomalous GVD, as well as the 3$\mathrm{^{rd}}$-order Kerr nonlinear optical effect.  When the pulse is focused by its own intensity, it is known as Self Phase Modulation (SPM); whereas when it is focused by a second pulse, it is referred to as Cross Phase Modulation (XPM) \cite{1,2}.  In addition to large GVD, PhCWG's can be fabricated out of highly 3$\mathrm{^{rd}}$-order-nonlinear dielectrics, dramatically reducing nonlinear length scales.  

The first such experimental measurement of soliton pulse compression in a PhCWG was conducted using Gallium Indium Phosphate (GaInP) PhCWG \cite{7}, which have the advantage of not having intensity-dependent two-photon absorption (TPA) and TPA-induced free-carrier absorption (FCA) \cite{9}.  The applications of this technology are numerous, as two-dimensional PhCWG may be fabricated into optical waveguides on a device the size of a computer chip, to be utilized for all-optical computing \cite{21}.  Optical computing has the advantage of significantly greater bandwidth compared to current solid-state devices, with reduced power and cooling, which can allow vastly improved computational performance \cite{22}.   

Another highly nonlinear dielectric material is silicon \cite{5,13}, which may also potentially be used to fabricate chip-scale optical waveguides.  Silicon is advantageous, as it is cheap, durable, readily available, and there already exists an infrastructure to mass-produce silicon devices.  Pulse compression was previously observed experimentally in novel silicon nanowires \cite{14,15}, but previous analytical and numerical studies have demonstrated that with SPM alone, TPA and FCA will prevent a pulse from self-focusing itself sufficiently to generate a soliton-like wave \cite{9}.  Even if the soliton pulse were to compress, it would simply dissipate after propagating relatively short nonlinear lengths.  One of the advantages of XPM is that a pulse can be focused without needing to be highly energetic and intense \cite{1}; therefore there was interest in seeing if a second pump pulse could be used to focus the primary signal pulse, to induce the soliton.  If XPM can be used to overcome the limitations of TPA and FCA and enable chip-scale soliton propagation in silicon, it may very well develop into practical all-optical computing.  

\section{Nonlinear Schr\"{o}dinger Equation}
In order to accurately model the optical soliton, two crucial nonlinear characteristics are necessary, the GVD and the nonlinear Kerr effect \cite{1}.  The GVD is determined from the derivatives of the refractive index over the angular frequency.  The other component of soliton pulse propagation is the nonlinear Kerr effect, which results in a change of refractive index proportional to the square of the electric field, or proportional to the optical intensity.  When there is strong SPM and/or XPM interplaying with strong anomalous GVD, the SPM/XPM will cause the leading edge frequencies to be lowered, and the trailing edge frequencies to be raised.  At the same time, the anomalous GVD will cause the lowered leading edge frequencies to slow down, and the raised trailing edge frequencies to speed up.  This results in the pulse compressing temporally into a soliton pulse \cite{1,2}.  

The GVD is determined by knowing the derivatives of the refractive index over the angular frequency.  These terms are found using the following equations \cite{1,2,4}:
\begin{eqnarray}
\beta(\omega)&=&n(\omega)\frac{\omega}{c_{0}}, \\ 
{\beta_{M}(\omega)}&=&\frac{d^{M}\beta(\omega)}{d \omega^{M}}, \\
\beta_{1}(\omega)&=&\frac{1}{v_{G}}=\frac{n_{G}}{c_{0}}=\frac{1}{c_{0}}(n(\omega)+\omega \frac{dn(\omega)}{d\omega}), \\
\beta_{3}(\omega)&=&\frac{d{\beta_{2}(\omega)}}{d\omega}= \frac{d^{2}{\beta_{1}(\omega)}}{d\omega^{2}}= \frac{d^{3}{\beta(\omega)}}{d\omega^{3}}, \\ \nonumber
\end{eqnarray}
where $n(\omega)$ is the refractive index as a function of angular velocity, $\omega$ ($\mathrm{{rad}/{s}}$) is the angular frequency, $c_{0}$ (m/s) is the speed of light, $v_{G}$ (m/s) is the group velocity, and $n_{G}$ is the group index.  These values can be determined either experimentally \cite{5,19}, approximated with the Sellmeier equation \cite{1,2}, or engineered to specific values with geometric dispersion, as is the case of photonic crystals \cite{6,8}.   

The other component of generating soliton pulse propagation and compression is the nonlinear Kerr effect.  The change in refractive index is determined by: 
\begin{eqnarray}
\Delta n &=& {n_2}{\cdot}({I_1}+{2{\cdot}{I_2}}), \\ 
I &=& \frac{{\vert}{A}{\vert}^2}{2{(\eta_0/n)}}, \\ \nonumber
\end{eqnarray}
where $n_2$ ($\mathrm{{m^2}/{W}}$) is the nonlinear Kerr coefficient, $\eta_0$ is the vacuum impedance ($\approx{120{\pi}}$ $\Omega$), $\emph{n}$ is the refractive index, \emph{A} (V/m) is the electric field, and $I_1$ and $I_2$ ($\mathrm{{W}/{m^2}}$) are, respectively, the intensities of the pulse and another coupled pulse.  This nonlinear term can be determined from the 3$\mathrm{^{rd}}$-order nonlinear susceptibility, which is also responsible for four-wave mixing \cite{2,20}: 
\begin{eqnarray}
{n_2} &=& {3{\eta_0}{\cdot}}{\frac{{\chi}^{(3)}}{{{\varepsilon}_0}{\cdot}n}}, \\ \nonumber
\end{eqnarray}
where $\epsilon_0$ is the electric permittivity in a vacuum (8.854 $\cdot$ 10$\mathrm{^{-12}}$ F/m), and $\chi^{(3)}$ $\mathrm{({{m^2}/{V^2}})}$ is the $\mathrm{3^{rd}}$-order nonlinear susceptibility coefficient \cite{2}.  Throughout this study, a Kerr nonlinearity of $\mathrm{4{\cdot}{10^{-18}}{\hspace{2 mm}}{m^2}/{W}}$ \cite{5,8,28} is used.  The nonlinear phase shift is thus \cite{1}: 
\begin{eqnarray}
{\Delta}{{\phi}_{NL}} &=& {n_2}{\cdot}{k_0}{\cdot}L{\cdot}({I_1}+{2{\cdot}{I_2}}) = {n_2}{\cdot}{\frac{2{\pi}}{\lambda}}{\cdot}L{\cdot}({I_1}+{2{\cdot}{I_2}}), \\ \nonumber
\end{eqnarray} 
where $k_0$ (m$\mathrm{^{-1}}$) is the optical wave-number, ${\lambda}$ (m) is the wavelength, and \emph{L} (m) is the optical propagation length.  These two nonlinear optical effects can be combined to form the NLSE, which in its complete form is represented as: 
\begin{eqnarray}
\frac{\partial A}{{\partial z}}= {\frac{i}{2}{\beta_{2}}{\frac{\partial A^{2}}{{\partial t^{2}}}}}+{\frac{i}{6}{\beta_{3}}{\frac{\partial A^{3}}{{\partial t^{3}}}}}+{{i}{\gamma{|A|^{2}}A{\cdot}{exp(-\alpha z)}}},\\
\gamma={\frac{2{\pi}{n_{2}}}{A_{e}{\cdot}{\lambda}}}{\cdot}{(\frac{n_{g}}{n})^{2}},
\end{eqnarray}
where $\beta_{2}$ ($\mathrm{s^2/m}$) is the $\mathrm{2^{nd}}$-order GVD coefficient, $\beta_{3}$ ($\mathrm{s^3/m}$) is the $\mathrm{3^{rd}}$-order GVD coefficient, $A_{e}$ $\mathrm{(m^2)}$ is the effective cross-sectional area of the waveguide, $\gamma$ $\mathrm{(m{\cdot}W)^{-1}}$ is the nonlinear parameter, $\alpha$ ($\mathrm{m^{-1}}$) is the linear loss coefficient, \emph{i} = $\sqrt{-1}$, and \emph{A} (V/m) is the pulse electric field envelope function.  \\

The NLSE is also known as the Split-Step Fourier method because each incremental distance step involves two fundamental steps.  The first step is to account for the GVD.  A code will convert the pulse information into the spectral domain by using the Fast Fourier Transforms (FFT) method \cite{1,4,26,27}, apply the GVD, and then convert the pulse back from the spectral domain to the temporal domain to calculate the Kerr nonlinearity.  The equation for the GVD is as follows \cite{1,4}: 
\begin{eqnarray}
{\bar{A}}(z,\omega) &=& {{\bar{A}}(0,\omega)}{\cdot}exp[({\frac{i}{2}}{\cdot}{{\beta}_2}{\cdot}{{\omega}^2}{\cdot}z)+({\frac{i}{6}}{\cdot}{{\beta}_3}{\cdot}{{\omega}^3}{\cdot}z)+{...}], \\ 
{\bar{A}}(z,\omega) &=& \int_{-\infty}^\infty {{A}(z,T)}{\cdot}{{exp}[-2{\pi}{i}{\omega}{T}]}\,\mathrm{d}T, \\ 
T &=& t - {\frac{z}{v_g}}, \\ 
\nonumber
\end{eqnarray}
where $\emph{A}(z,T)$ and $\bar{A}(z,\omega)$ are the pulse envelope functions in the temporal and spectral domains, $\emph{T}$ is the normalized time, $v_g$ is the group velocity, $\emph{z}$ is the propagation distance, and $\omega$ is the normalized angular frequency.  The second step is to take the pulse, now in the temporal domain, and apply the effects of SPM/XPM.  The equation for the nonlinear phase shift is \cite{1,4}: 
\begin{eqnarray}
{{A_1}(z,T)} &=& {{{A_1}}(0,T)}{\cdot}exp[i{\cdot}{n_2}{\cdot}{\frac{2{\pi}}{\lambda}}{\cdot}L{\cdot}{\frac{({{\vert}{A_1(0,T)}{\vert}^2}+{2{\cdot}{{\vert}{A_2(0,T)}{\vert}^2}})}{2({\eta_0}/{n})}}], \\ \nonumber
\end{eqnarray}
After completely these two steps, one can apply the effects of linear and nonlinear loss for the specified distance increment.  By following this algorithm, soliton pulse propagation can be simulated and studied.  

\section{Two Photon Absorption}
One big challenge to soliton pulse propgation in silicon is the nonlinear effect of multi-photon absorption \cite{9}.  Dielectric materials have an electronic band gap, where photons with energy greater than the electronic band gap get absorbed by the material \cite{10}.  The electronic band gap for silicon is 1.1 electron-volts (eV) \cite{7,9}.  As it is desired to use light pulses at a wavelength of 1550 nm, this corresponds to a photon energy of 0.8 eV, and thus silicon is transparent at this wavelength.  Multi-photon absorption, however, is the result of multiple photons interacting, and their combined energy exceeding the electronic band gap.  For example, at 1550 nm, two photons will result in a total energy of 0.8$\times$2=1.6 eV, which exceeds the 1.1 eV electronic band gap of silicon.  Therefore, silicon is subject to TPA at 1550 nm.  

In order to calculate the impacts of TPA on silicon, one must find the coefficient of TPA at the given wavelength by the following equations \cite{11}: 
\begin{eqnarray}
{{\beta}_{TPA}}(\omega) &=& K{\cdot}{\sqrt{\frac{EP}{{{n_0}^2}{\cdot}{{E_g}^3}}}}{\cdot}{F_2}(\frac{2{\pi}{\hbar}{\cdot}{\nu}}{E_g}), \\ 
{F_2}(x) &=& \frac{{({{2x}-1})}^{\frac{3}{2}}}{{(2x)}^5}, \\ 
{2{\pi}{\hbar}{\cdot}{\nu}} &=& {\frac{1240}{\lambda(nm)}}{\indent}(eV), \\ \nonumber
\end{eqnarray}
where $\emph{EP}$ is equal to 21, $\hbar$ is Plank's constant over $2{\pi}$, $\nu$ is the frequency, $n_0$ is the refractive index, and $E_g$ is equal to the 1.1 eV electronic band-gap of silicon \cite{7,9,13}.  The constant $\emph{K}$ was calculated based on the experimentally measured value of $4.4{\cdot}{10^{-12}}$ (m/W) at 1550 nm \cite{12}.  With the value of ${\beta}_{TPA}$, the TPA can be simulated with \cite{13,25}: 
\begin{eqnarray}
{I(z,t)} &=& {I(0,t)}{\cdot}{exp[{-}{{\beta}_{TPA}}{\cdot}{I(0,t)}{\cdot}{z}]}, \\ \nonumber
\end{eqnarray}

Once the two-photon absorption coefficient is determined, one can determine the impact from the free carriers generated by the TPA.  For ultrashort pulses, where the pulse is substantially shorter than the free-carrier lifetime, the FCA tends to be substantially less than the effects of TPA \cite{13}.  The model calculates the FCA at each distance step separately, assuming that there were no free-carriers there at the beginning.  This is an approximation, as carriers will in fact drift over to the different distance steps, but this is difficult to accurately predict.  For this reason, the assumption that carriers stay localized will be used.  After the SPM/XPM and dispersion was applied to the pulse, the model predicts the rate of free-carrier generation from the two-photon absorption \cite{12,13,23,25},  
\begin{eqnarray}
\frac{dN(t)}{dt} &=& G - {\frac{N(t)}{{\tau}_{FCA}}}, \\
G &=& {\frac{{{\beta}_{TPA}}{\cdot}{I^2}}{4{\pi}{\hbar}{\nu}}}, \\
\nonumber
\end{eqnarray}
where ${\beta}_{TPA}$ (m/W) is the coefficient of two photon absorption, $\emph{I}$ ($\mathrm{W/m^2}$) is the optical intensity (either from the signal or coupled pulse), $\nu$ is the frequency, ${\hbar}$ is Plank’s Constant over 2$\pi$, $\emph{G}$ is the rate of free-carrier generation, ${\tau}_{TPA}$ (s) is the free carrier lifetime, and $\emph{N(t)}$ is the free-carrier density.  Once the rate of free-carrier generation over time has been determined at the specific distance step, the next step is to calculate the density of free-carriers.  While the simulation goes through the process of generating carriers over time, it also reduces carriers based on the specified free-carrier lifetime.  After the carrier density is determined as a function of time (at the given distance increment), the appropriate free-carrier absorption is applied to the pulse in the temporal domain.  The equation is as follows \cite{13,25}: 
\begin{eqnarray}
I(z)={I(0)}{\cdot}exp[{-{\frac{1}{2}}}{{\cdot}(1.45{\cdot}{10^{-21}}){{\cdot}N(t){\cdot}z}}], \\
\nonumber
\end{eqnarray}
By implementing these equations within the NLSE simulation, an accurate analysis of soliton pulse propagation in a silicon waveguide subjected to TPA and FCA can be achieved.  

\section{Simulations of Cross Phase Modulation}
The NLSE model that will be used throughout this study was first set to a single pulse with SPM.  First, the model propagated a 10 pJ hyperbolic secant squared pulse down a 10 cm silicon nanowire; the results of this simulation are shown in Figure 1-a.  A compression ratio factor of 7 was clearly observed.  The model then scaled the pulse energy and waveguide length to match the soliton number \cite{7} and nonlinear length \cite{1} as the nanowire study, using the following relations:
\begin{eqnarray}
L'={L}{\cdot}{|{\frac{{\beta}_2}{{\beta}'_2}}|}, \\
E'={E}{\cdot}{|{\frac{{\beta}'_2}{{\beta}_2}}|}{\cdot}{[{\frac{A'_e}{A_e}}]}{\cdot}{[{\frac{{n_G}{\cdot}{n'}}{{n'_G}{\cdot}{n}}}]}, \\
\nonumber
\end{eqnarray}
where $\emph{L}$ (m) is the waveguide length, $\emph{E}$ (J) is the input pulse energy, ${\beta_2}$ $\mathrm{(s^2/m)}$ is the 2$\mathrm{^{nd}}$-order GVD coefficient, ${A_{e}}$ $\mathrm{(m^2)}$ is the waveguide effective cross-sectional area, and $n_{G}$ and $\emph{n}$ are the group and refractive indices, respectively.  Based on these relations, it is clear that by using a PhCWG, the soliton length scales can become reduced dramatically.  The caveat is that it is necessary to use higher pulse-energies, which in turn have greater TPA and FCA, which attenuate the soliton, and disturb the pulse, as observed in Figure 1b.  While compression has been observed at the early onset of soliton propagation within the PhCWG, these compressed pulses simply cannot sustain themselves in the presence of intense TPA and FCA.
\begin{figure}[H]
\centering
\begin{minipage}[b]{0.45\linewidth}
\centering
\includegraphics[width=\textwidth]{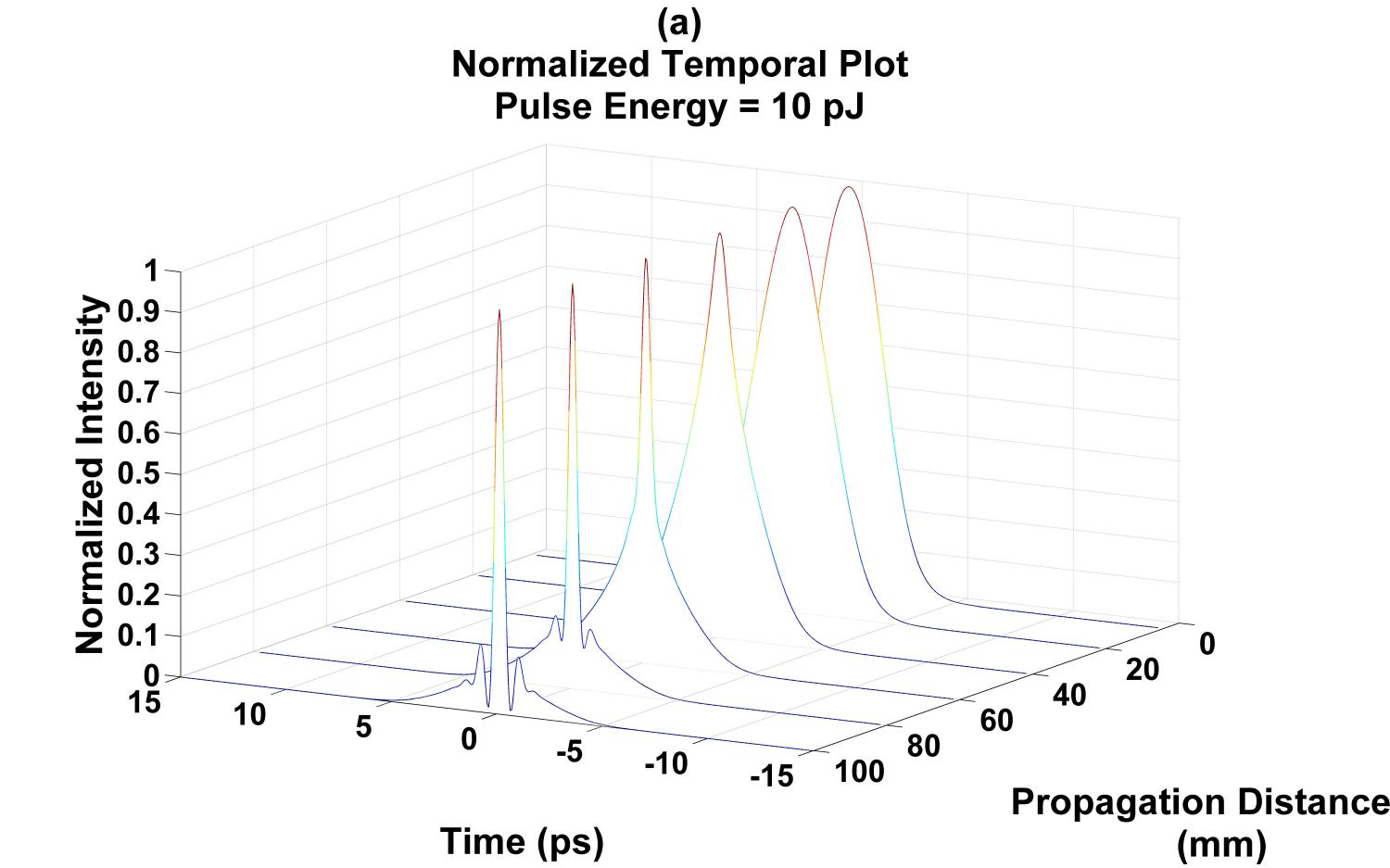}
\label{fig:1a}
\end{minipage}
\begin{minipage}[b]{0.45\linewidth}
\centering
\includegraphics[width=\textwidth]{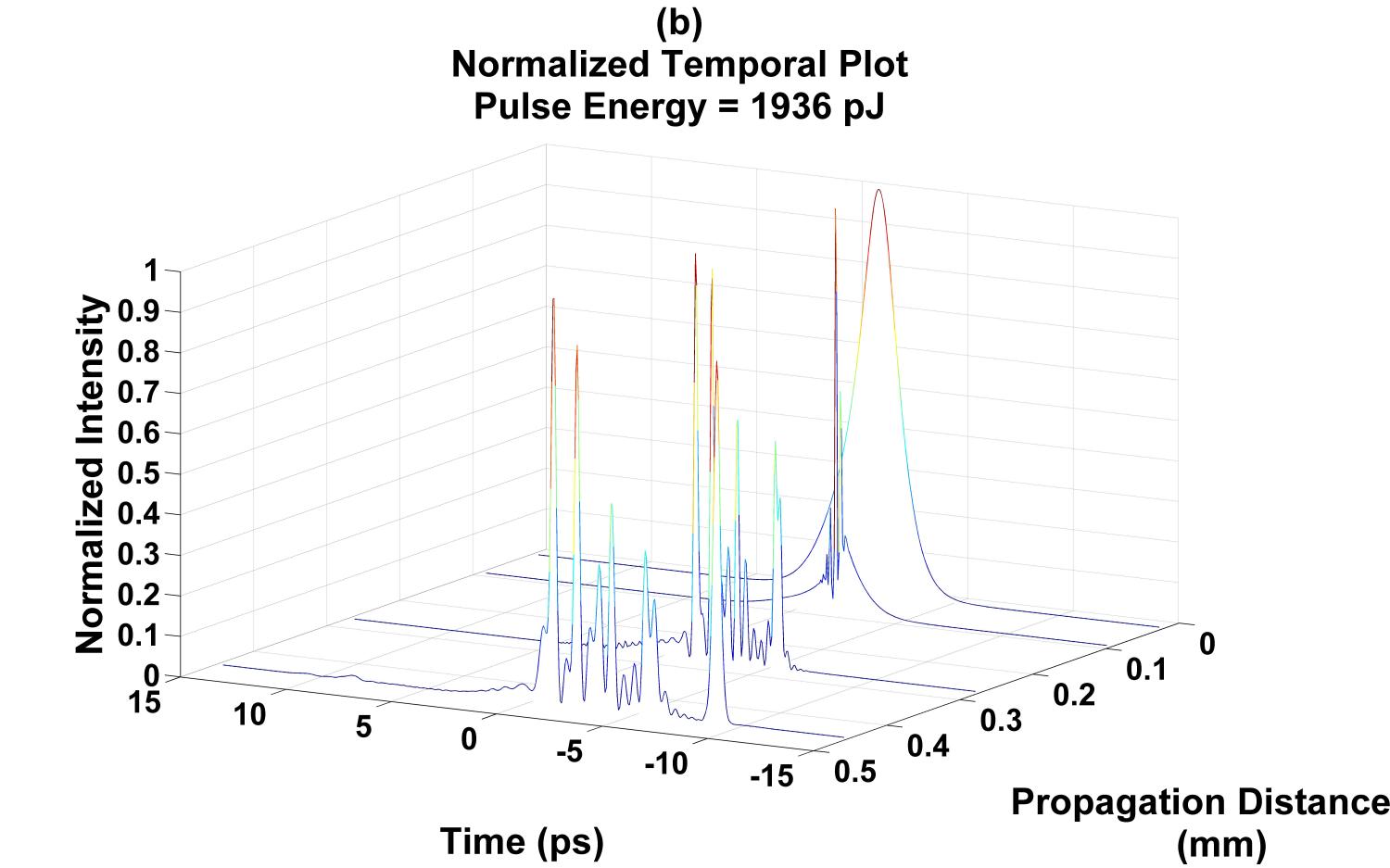}
\label{fig:1b}
\end{minipage}
\caption{SPM-induced soliton propagation, in a (a) silicon nanowire and a (b) silicon PhCWG.}
\label{fig:fig1}
\end{figure}

The NLSE model was modified to study both XPM as well as the effects of cross-TPA (XTPA) and the resulting FCA \cite{13}.  In this XPM simulation the process is completed twice, first for the primary pulse, and second for the secondary pulse.  It uses a modified split-step, since there are two coupled pulses, to take into account differences in the group velocity which result from the different wavelengths between the pump and the signal.  

Part of the modified aspect is a loop for the nonlinear effects that require considering both pulses, the two include XPM, XTPA, and FCA induced from XTPA.  The split-step calculations are used in a normalized time domain \cite{1}, however, with different wavelengths and thus different group velocities, this normalized time is not always applicable.  In fact, one concern with XPM is the case of walk-off where a faster traveling pulse fully clears away from the slower pulse within the waveguide \cite{16}.  A time shift factor is determined, based on the difference in group velocities, defined as ${\delta}$T, where
\begin{eqnarray}
{\delta}T&=&{({n'_G}-{n_G})}{\cdot}{(\frac{{\delta}z}{{\delta}t{\cdot}{c_0}})}, \\
\nonumber
\end{eqnarray}
where ${\delta}z$ is the distance increment, ${\delta}t$ is the time step used in the simulation, and $c_{0}$ is the speed of light in a vacuum.  The value of ${\delta}T$ is rounded down, as it is an integer number of time steps to shift, and added to calculate the effects of pulse 2 onto pulse 1 (and subtracted for pulse 1 onto pulse 2).  

\section{Soliton Propagation with Cross Phase Modulation}
This model was run using the parameters experimentally observed in silicon PhCWG \cite{8}, with a negligibly low pulse energy (to avoid SPM) of 10 fJ pulse for the signal pulse near the highly-dispersive photonic band-gap wavelength at 1544 nm, coupled temporally with an intense 105 pJ pump pulse at 1536 nm.  In this particular example, the GVD at the signal wavelength was a highly dispersive anomalous -4 ps$\mathrm{^{2}}$/mm, whereas the pump wavelength had a normal GVD coefficient of 100 ps$\mathrm{^{2}}$/m.  The signal was also set to a third-order GVD coefficient of 0.1 ps$\mathrm{^{3}}$/mm, a value typical of these engineered photonic crystal waveguides \cite{7}.  While higher-order dispersion is known to affect soliton propagation, especially at wavelengths very close to the highly dispersive photonic crystal band-gap \cite{29}, there was negligible difference in outcomes for the practical lengths modeled in this study.  Finally, a linear loss of 2.4 dB/mm was added to the simulation.  \\

By the phenomena of XPM, temporal compression was observed; the signal pulse underwent compression by a factor of 5 after propagating through 120 $\mathrm{\mu}$m of the PhCWG.  This result is displayed both as a function of normalized time and propagation distance in Figure 2-a; as well as a comparison of the normalized temporal intensity function in Figure 2-b.  Finally, the pump pulse did not experience any compression, as it was subjected to much less GVD; this is demonstrated in Figure 2-c.  As a result of this, the simulation clearly demonstrates that the compression is the result of XPM, as no compression was observed when the same 10 fJ signal pulse propagates through the waveguide without the pump pulse. \\
\begin{figure}[H]
\centering
\begin{minipage}[b]{0.30\linewidth}
\centering
\includegraphics[width=\textwidth]{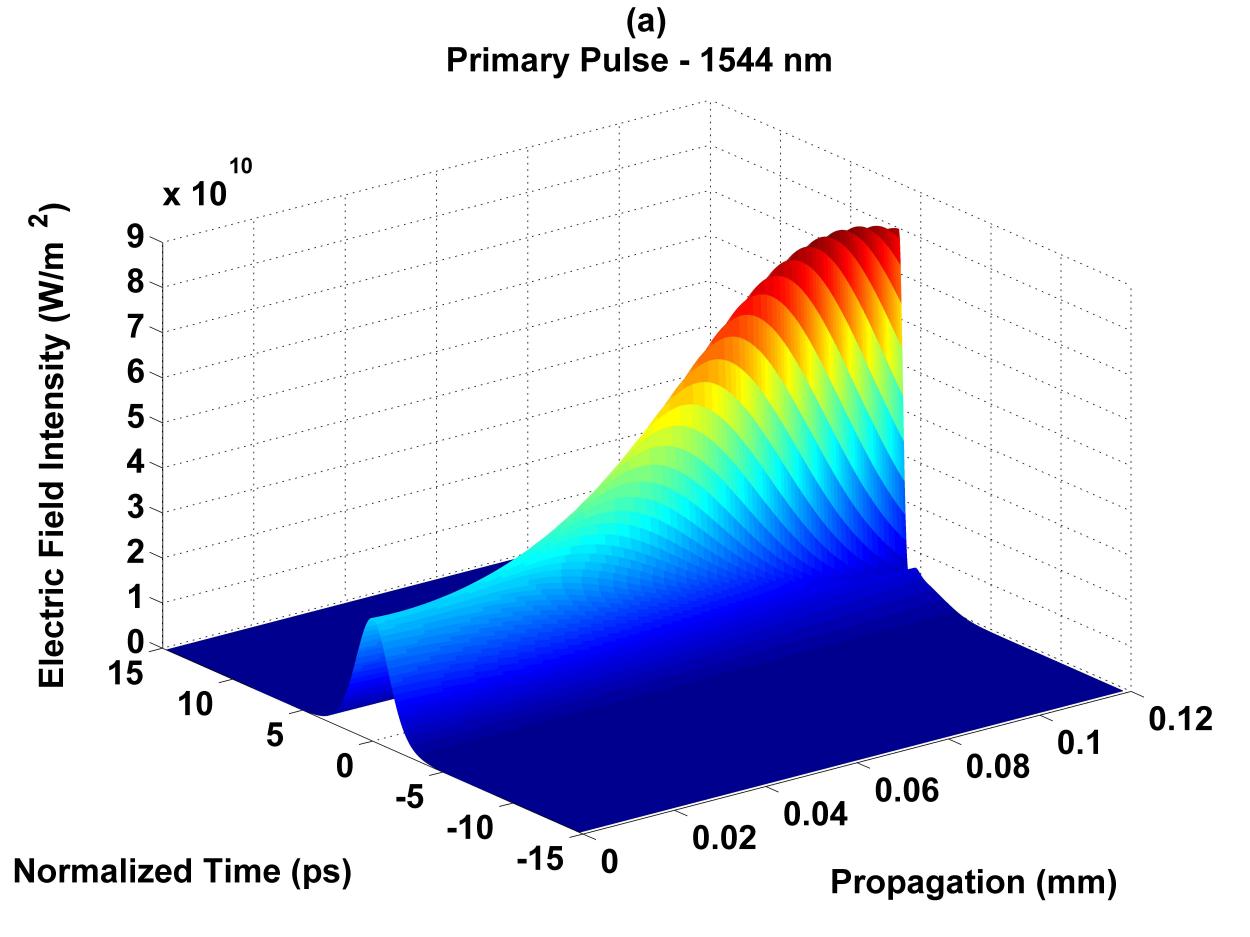}
\end{minipage}
\begin{minipage}[b]{0.30\linewidth}
\centering
\includegraphics[width=\textwidth]{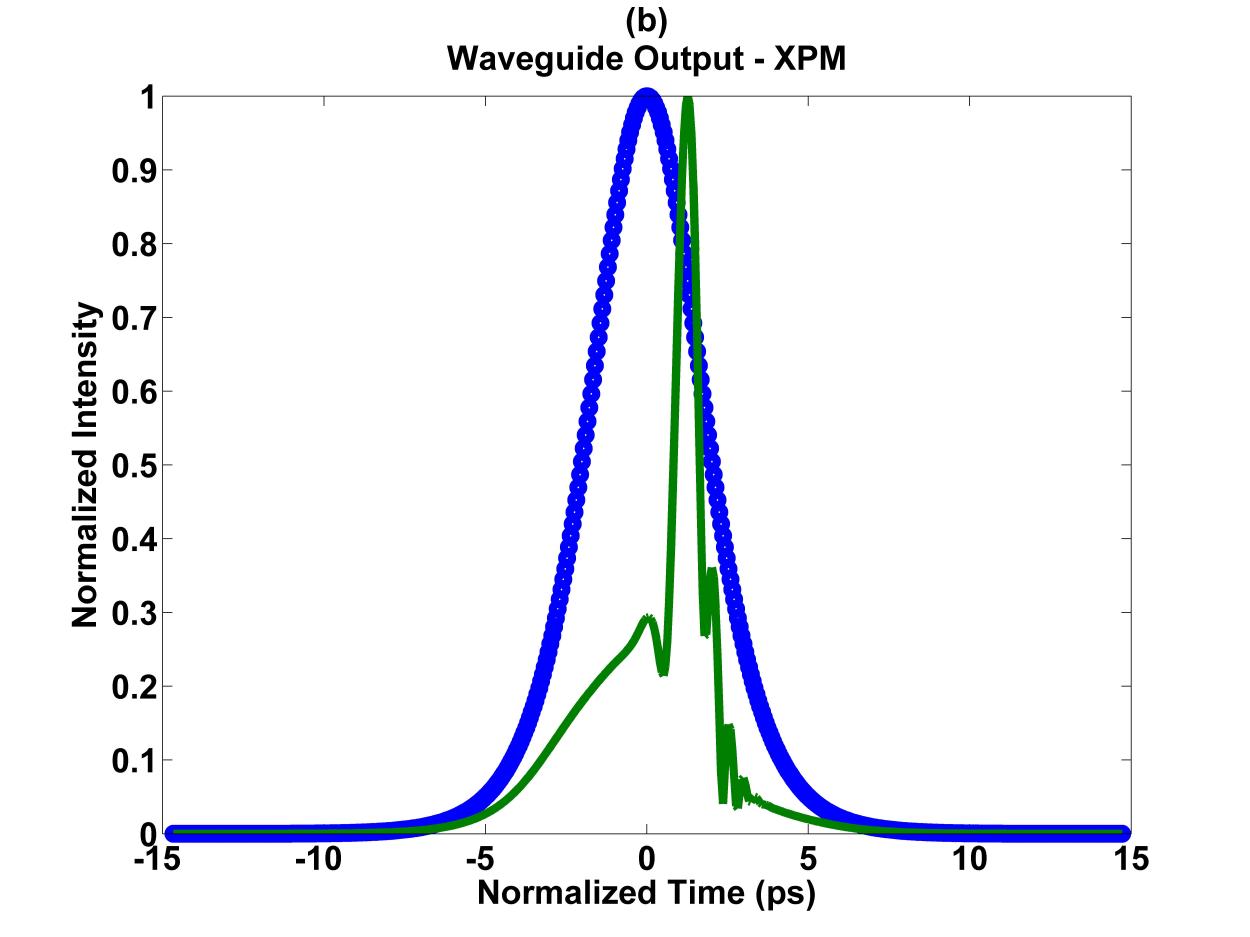}
\end{minipage}
\begin{minipage}[b]{0.30\linewidth}
\centering
\includegraphics[width=\textwidth]{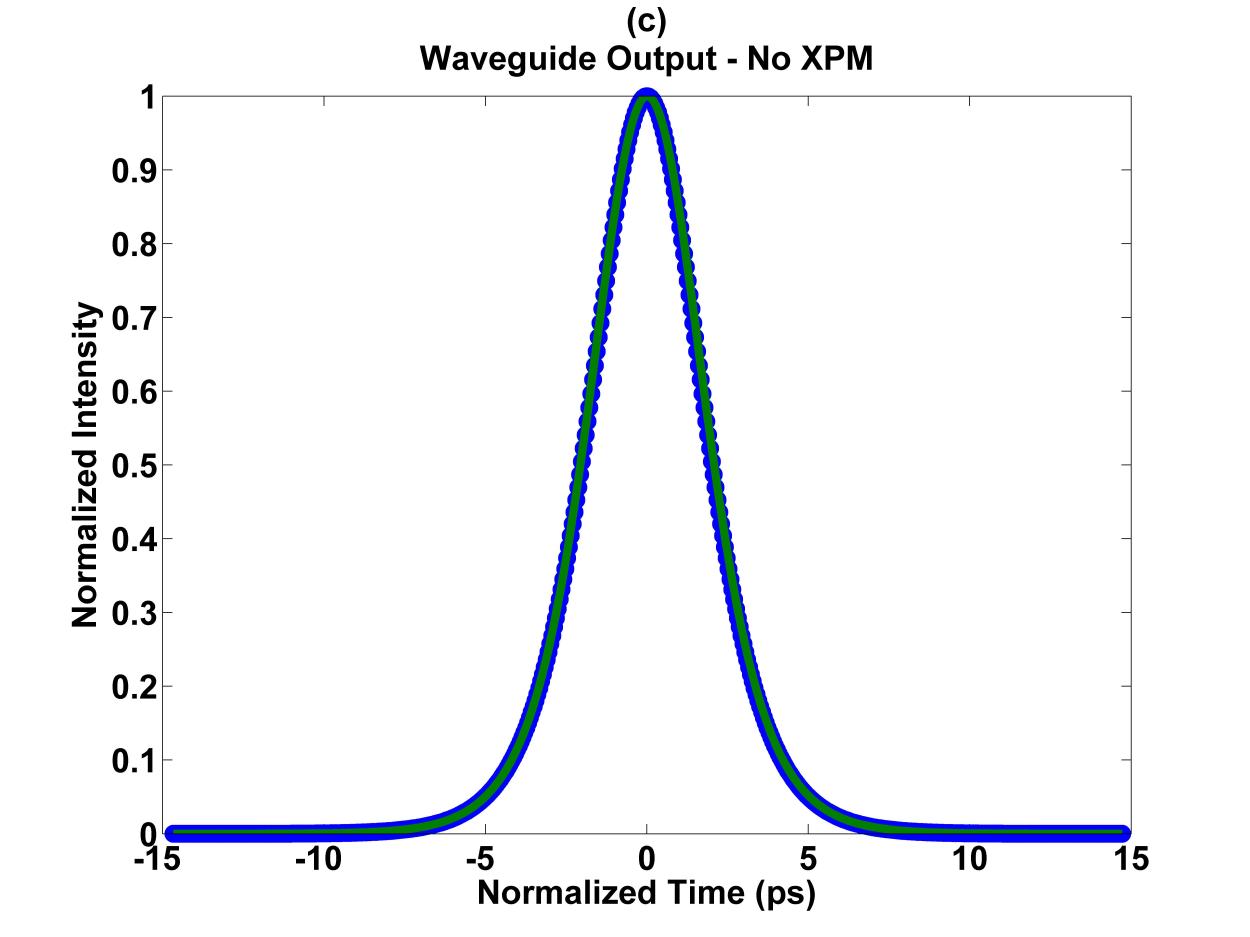}
\end{minipage}
\caption{XPM induced compression: (a) is the signal pulse compression as function of waveguide propagation distance and time; (b) is the waveguide output of weak signal pulse under XPM; and (c) is the waveguide output of weak signal pulse without XPM, thus proving the compression is the result of XPM.}
\label{fig:fig2}
\end{figure}

When running the simulation, a numerical scale was developed to represent XPM-induced solitons.  Much like a SPM-soliton \cite{7,9}, the XPM-soliton number $\emph{N}$ is a ratio of the fundamental soliton energy: 
\begin{eqnarray}
{N}&=&{\sqrt{\frac{E_P}{E_f}}}, \\
{E_f}&=&{{\frac{2|{\beta_2}|}{|{\gamma}|{\cdot}{\tau}}}}, \\ 
\nonumber
\end{eqnarray}
where ${\beta}_{2}$ $\mathrm{(s^2/m)}$ is the 2$\mathrm{^{nd}}$-order GVD coefficient, ${\gamma}$ $\mathrm{(W{\cdot}m)^{-1}}$ is the nonlinear Kerr coefficient, ${\tau}$ (s) is the temporal duration of the signal pulse in seconds, and $E_{p}$ (J) is the pump pulse energy.  The pulse compression ratio requires higher XPM soliton numbers than with SPM; this is expected due to the effect of pulse walk-off, which becomes more profound at longer waveguide lengths.  At the practical length scale of 0.1 nonlinear lengths (similar length studied with SPM), which equated to 0.4 mm (a typical photonic crystal waveguide \cite{5,7,8,9}), a pulse compression of nearly 3 was observed from an XPM-soliton number of 15; the results of these studies can be seen in Figure 3-a.  The Time Bandwidth Product (TBP) was tabulated and is represented by Figure 3-b.  Under the practical conditions mentioned earlier, the TBP remained relatively unchanged; evidence of XPM-induced soliton pulse propagation.  At longer length scales, the pulse eventually underwent broadening and the TBP increased; this can be attributed to TPB attenuating the pump pulse, as well as pulse walk-off.  The nonlinear loss (in dB) was also observed in figure 3-c.  In all cases, the longer the nonlinear propagation length, the more loss was observed; this is to be expected.  It was interesting to note, that in shorter propagation lengths, higher XPM-Soliton numbers had an ability to actually \emph{increase} the peak energy as the signal pulse was compressed by XPM.  In all cases studied, the optical pulses were all by-definition dissipative solitons; it was demonstrated that XPM can sustain the pulse over much longer propagation lengths scales than SPM alone.  
\begin{figure}[H]
\centering
\begin{minipage}[b]{0.30\linewidth}
\centering
\includegraphics[width=\textwidth]{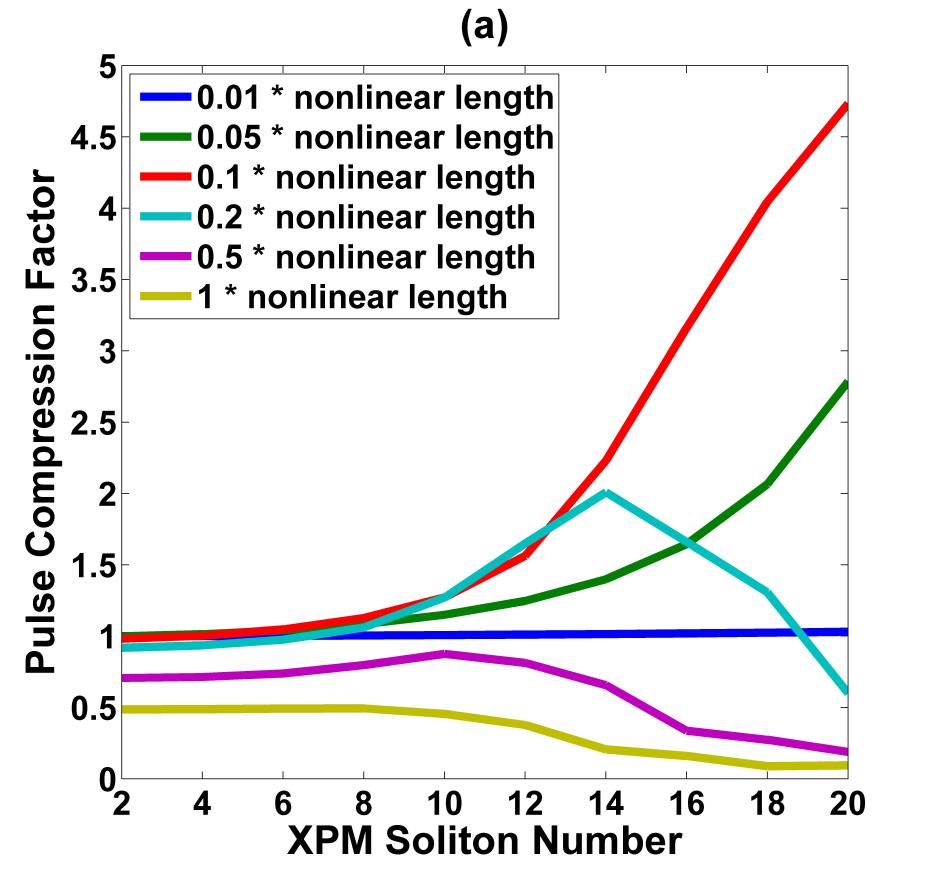}
\end{minipage}
\begin{minipage}[b]{0.30\linewidth}
\centering
\includegraphics[width=\textwidth]{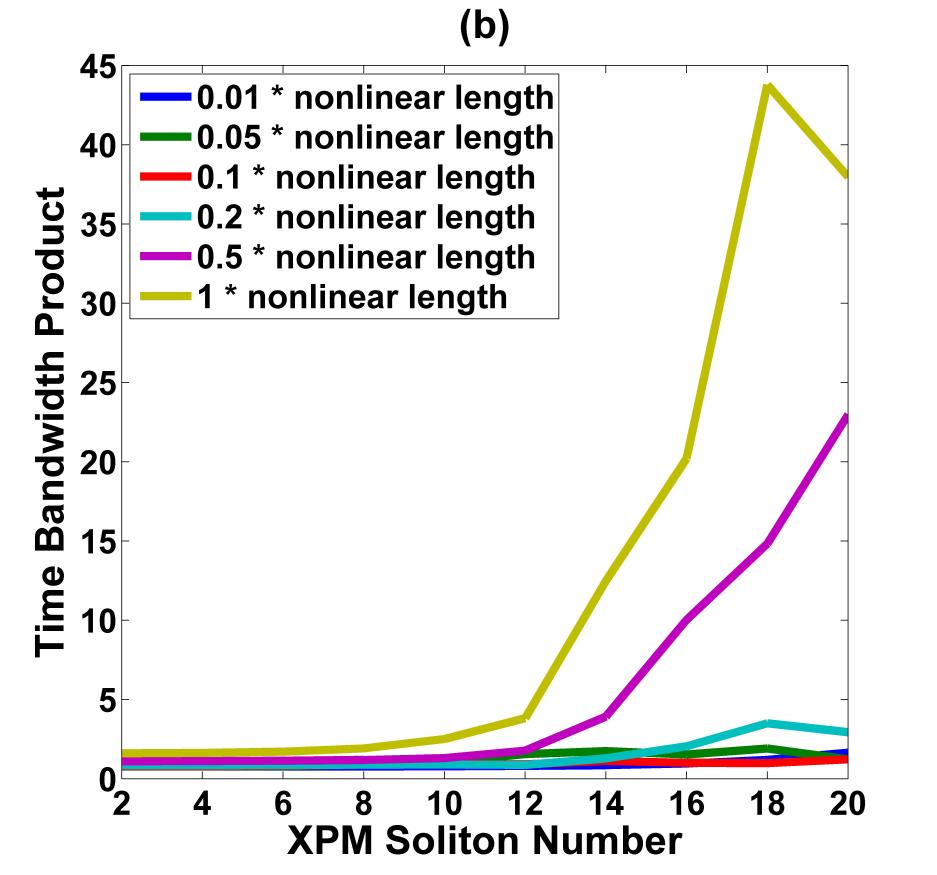}
\end{minipage}
\begin{minipage}[b]{0.30\linewidth}
\centering
\includegraphics[width=\textwidth]{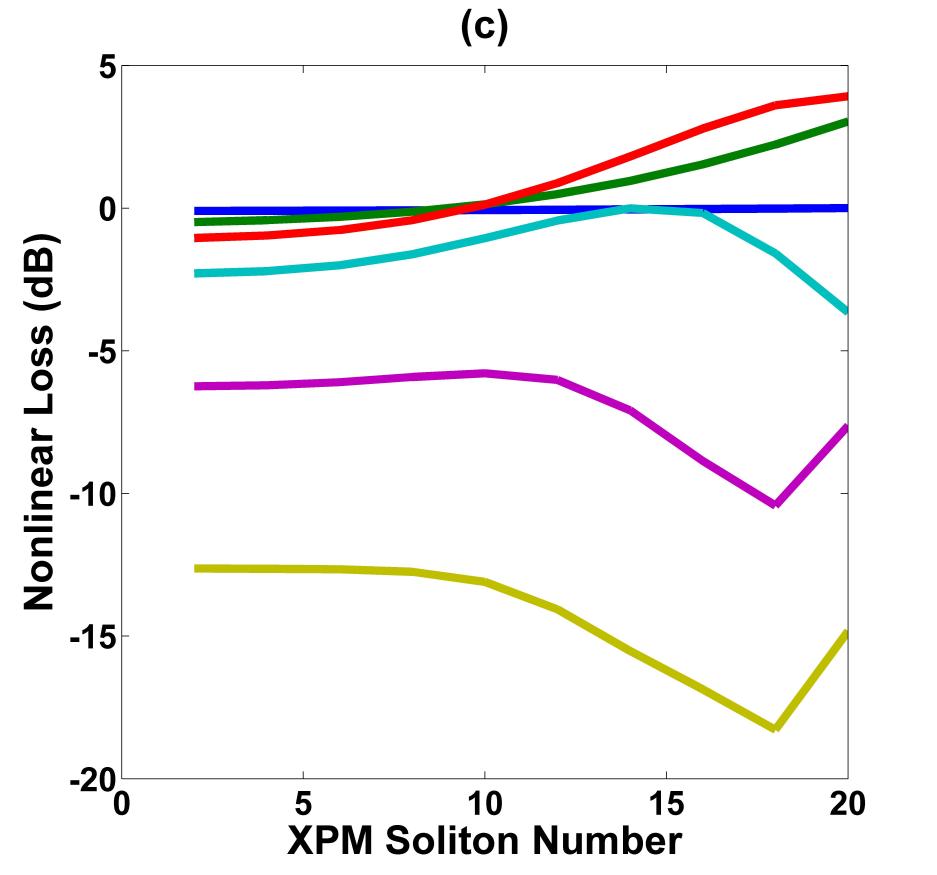}
\end{minipage}
\caption{Results of numerical XPM studies: (a) the signal pulse compression factor, (b) the output signal pulse TBP, and (c) the nonlinear peak intensity loss (dB).}
\label{fig:fig3}
\end{figure} 

\section{Practical Applications of Cross Phase Modulation}
The previous study was on the assumption that both the pump and signal wavelengths were subjected to TPA and FCA; this may be practically realized by utilizing a spectrally broad pulse and narrow bandpass filters \cite{16}.  While this may be simple to implement, it has the limitation of significant power attenuation from the filters.  Another potential option to generate optical pulses of different wavelengths coupled temporally and spatially is to use an Optical Parametric Oscillator (OPO), where a broad range of wavelengths may be generated by adjusting the length of the resonating cavity \cite{17,18}.  In an OPO, signal and idler pulses are generated with a combined wavelength energy equal to the pump wavelength energy: 
\begin{eqnarray}
\frac{1}{{\lambda}_{pump}} = {\frac{1}{{\lambda}_{signal}}} + {\frac{1}{{\lambda}_{idler}}}. \\
\nonumber
\end{eqnarray}
By utilizing this technology, one may realistically generate pulses coupled temporally and spectrally, where the pump pulse (not to be confused with the pump pulse for the OPO) has a wavelength not subjected to TPA within silicon.  In this next study, the authors investigate pulse compression resulting from a low-powered signal pulse at 1550 nm; and a pump pulse at a wavelength of 2300 nm, which is not subjected to TPA due to silicon's electronic band gap.  This can be achieved with a properly aligned OPO pumped 926 nm, a realistic wavelength that can be achieved in a titanium sapphire laser \cite{24}.  \\
\begin{figure}[H]
\centering
\begin{minipage}[b]{0.30\linewidth}
\centering
\includegraphics[width=\textwidth]{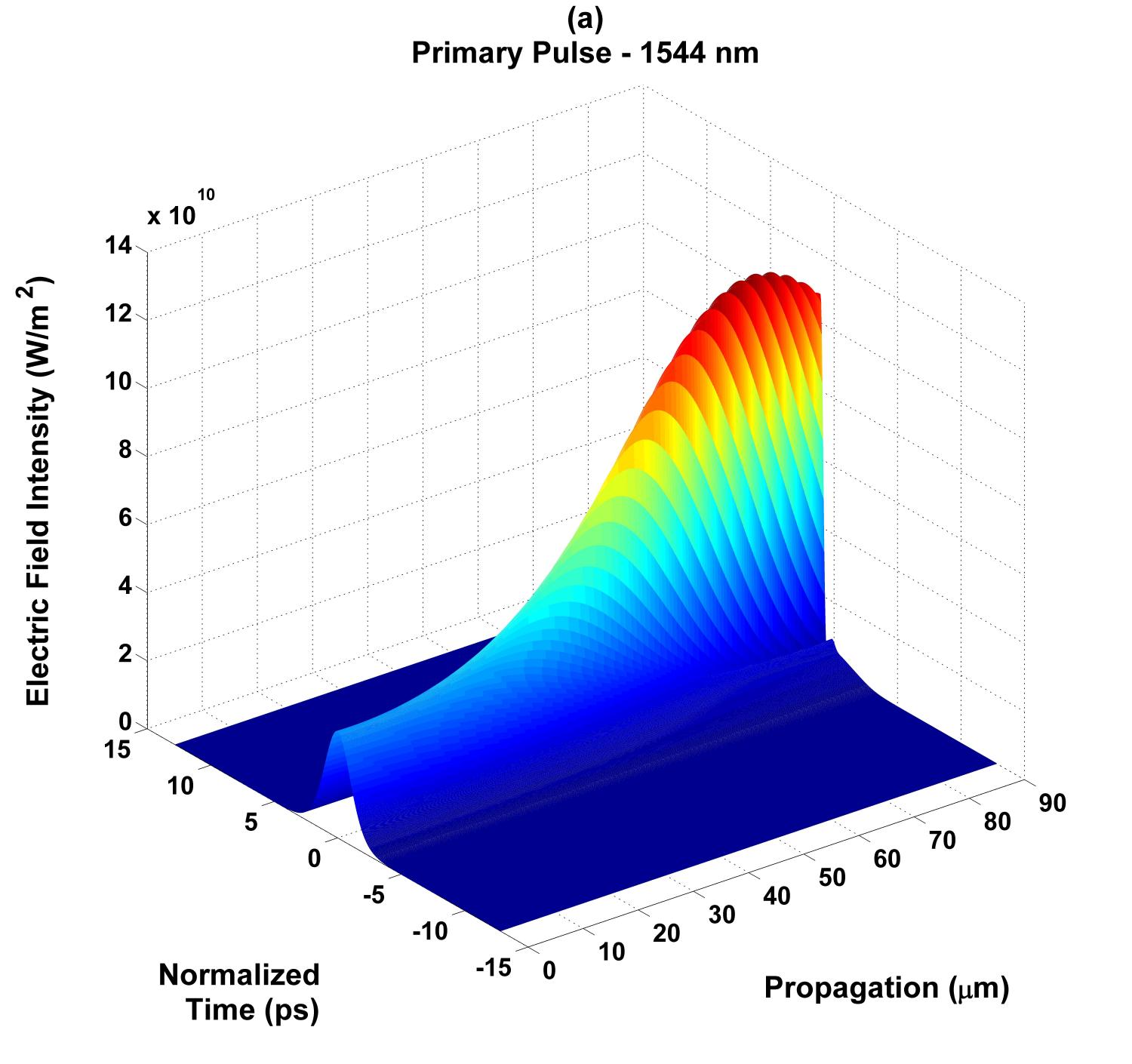}
\end{minipage}
\begin{minipage}[b]{0.30\linewidth}
\centering
\includegraphics[width=\textwidth]{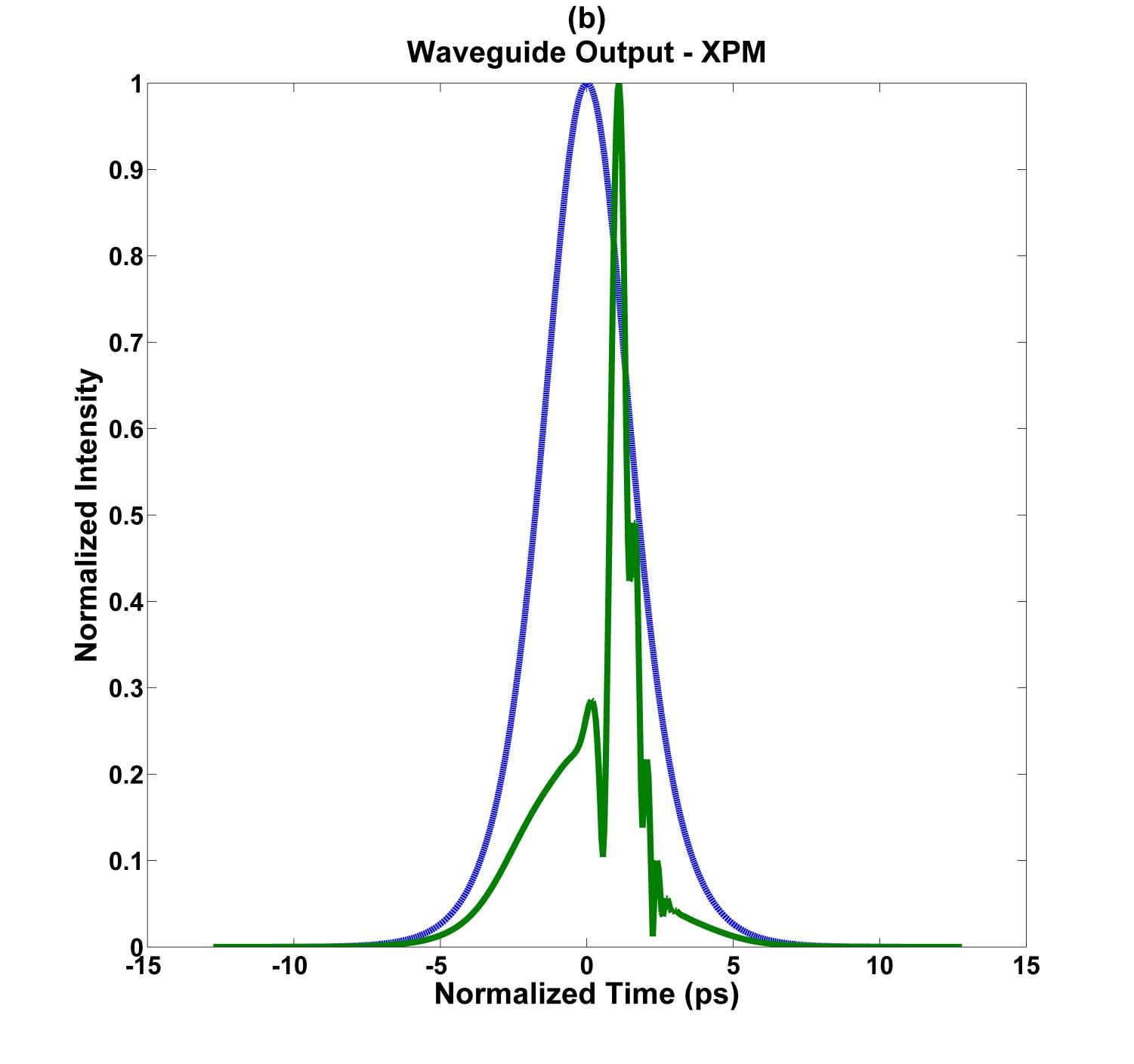}
\end{minipage}
\begin{minipage}[b]{0.30\linewidth}
\centering
\includegraphics[width=\textwidth]{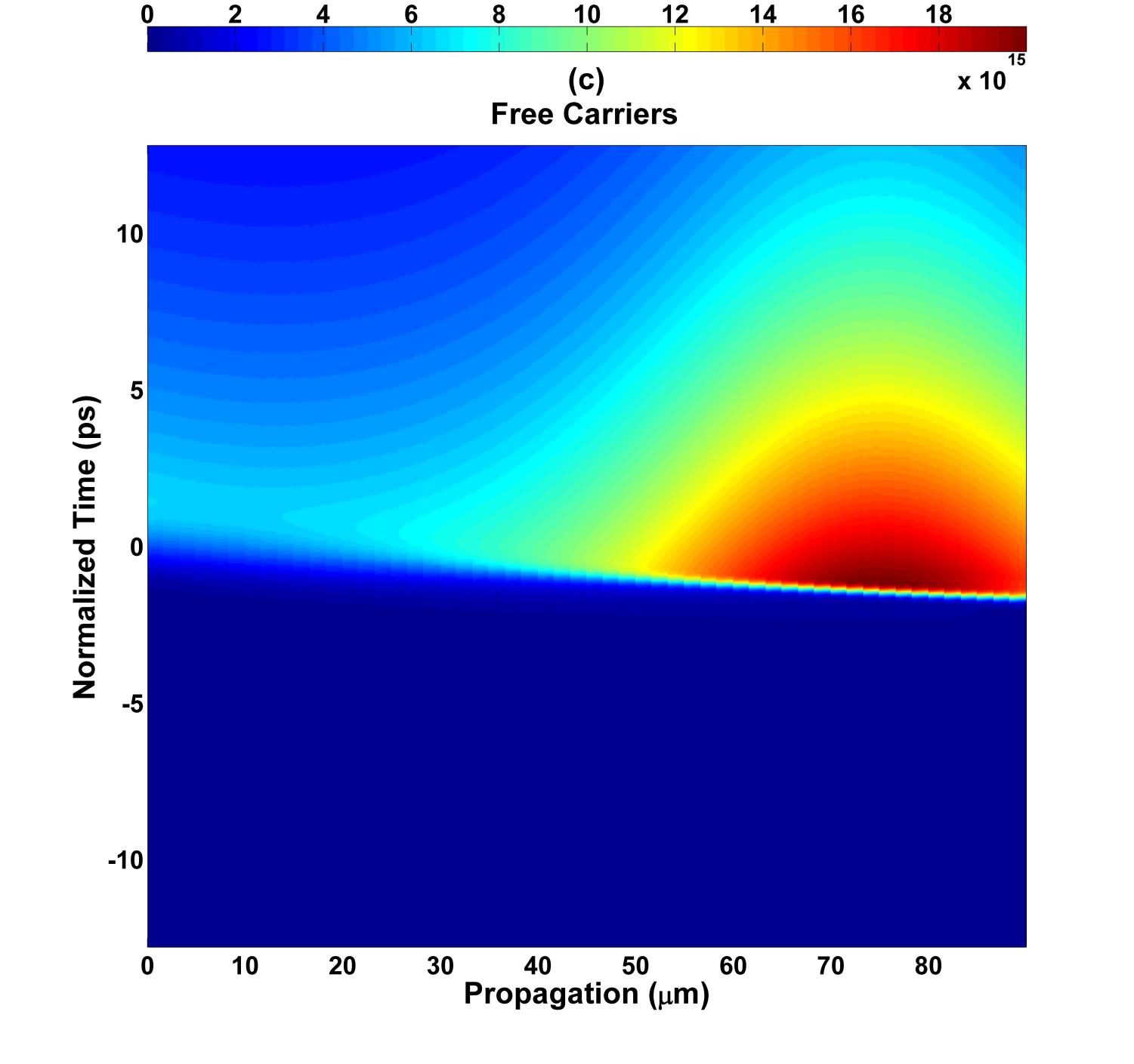}
\end{minipage}
\caption{Results of numerical XPM studies, with a low-energy C-band pulse coupled with a high-energy 2300 nm \emph{pump} pulse: (a) signal pulse compression as function of waveguide propagation distance and time; (b) numerical predicted free-carrier density as a function of waveguide propagation distance and time; and (c) waveguide output of weak signal pulse under XPM.}
\label{fig:fig4}
\end{figure} 

The study was conducted with the same dispersion and energy parameters as before, but without the pump pulse subjected to TPA, and a compression ratio over 5 was observed at 90 $\mathrm{\mu}$m of propagation.  This is shorter than the 120 $\mathrm{\mu}$m of propagation needed for the study in Figure 2; and the results of this study can be observed both in Figure 4-a and Figure 4-b.  It was interesting to observe that the generation of free carriers, as shown in Figure 4-c, actually increased as the signal pulse, which is subjected to TPA, compressed and increased in temporal intensity due to the XPM from the pump pulse.  In addition, it was observed that the compression remained consistent for even the longer propagation studies.  This study was repeated for a host of different XPM Soliton Numbers (Figure 5-a) and nonlinear length ratios (Figure 5-b), and the nonlinear losses were observed (Figure 5-c).  These nonlinear losses were very close to the previous non-OPO study (Figure 3-c).  The data clearly shows that by utilizing a substantially longer wavelength pulse, practical soliton pulse propagation in silicon can be both achieved and sustained for low-energy pulses at wavelengths subjected to intensity-dependent TPA.  \\

\begin{figure}[H]
\centering
\begin{minipage}[b]{0.30\linewidth}
\centering
\includegraphics[width=\textwidth]{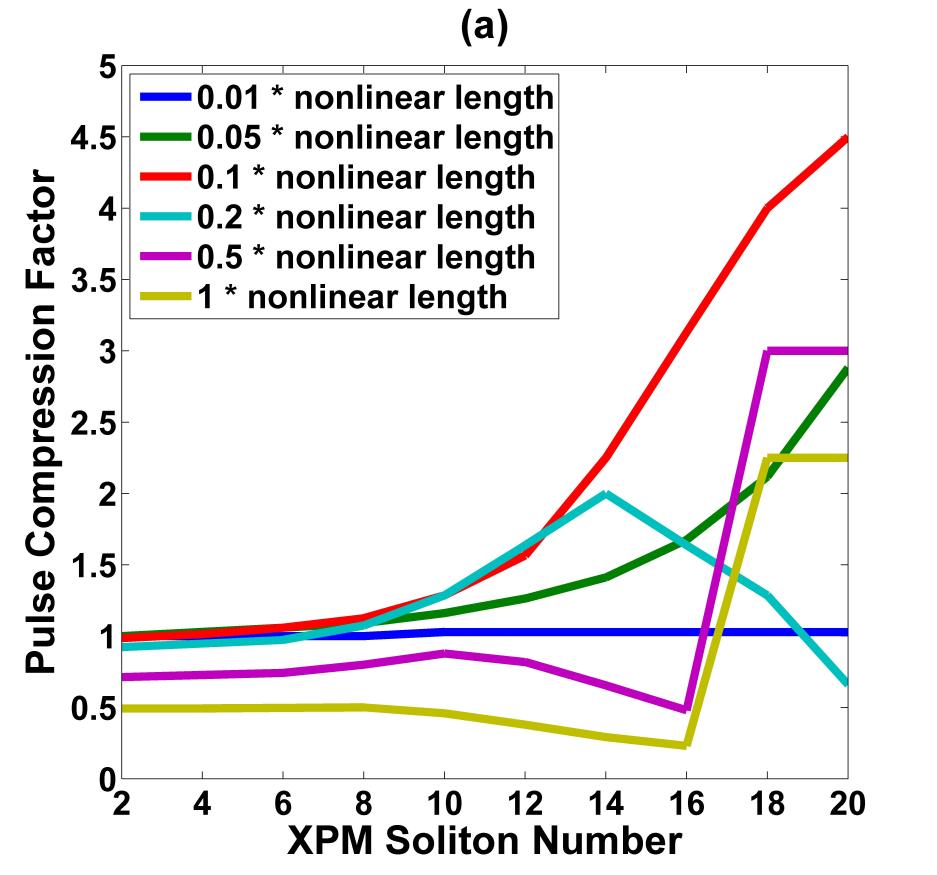}
\end{minipage}
\begin{minipage}[b]{0.30\linewidth}
\centering
\includegraphics[width=\textwidth]{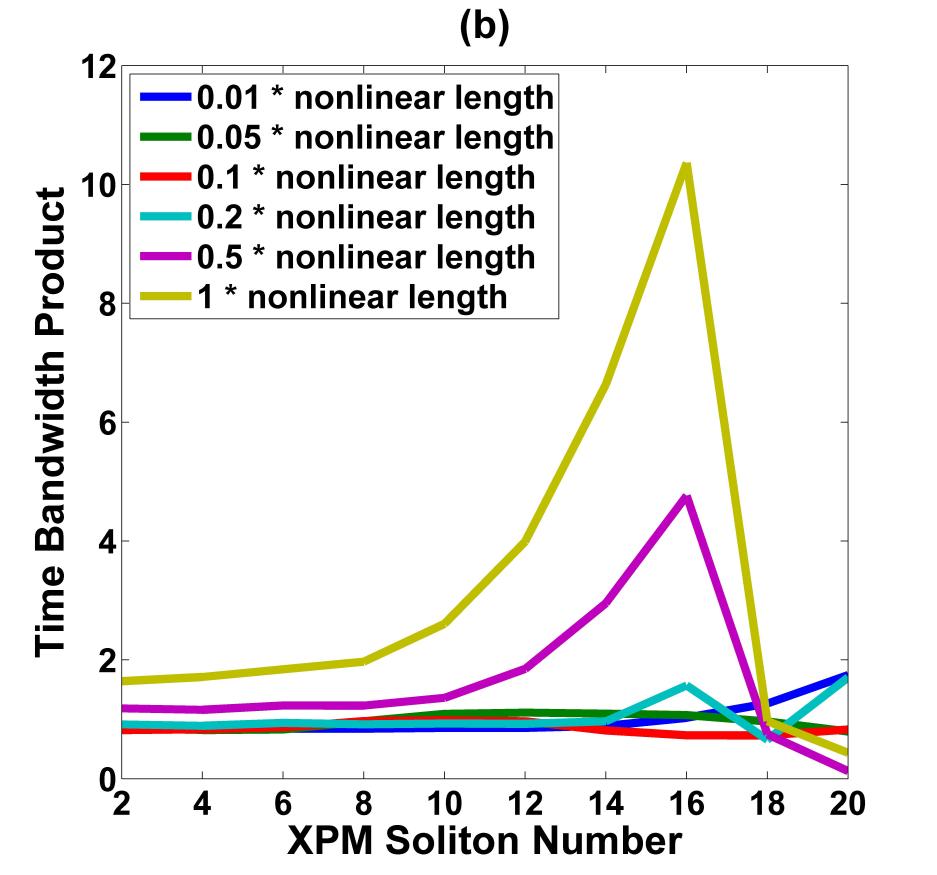}
\end{minipage}
\begin{minipage}[b]{0.30\linewidth}
\centering
\includegraphics[width=\textwidth]{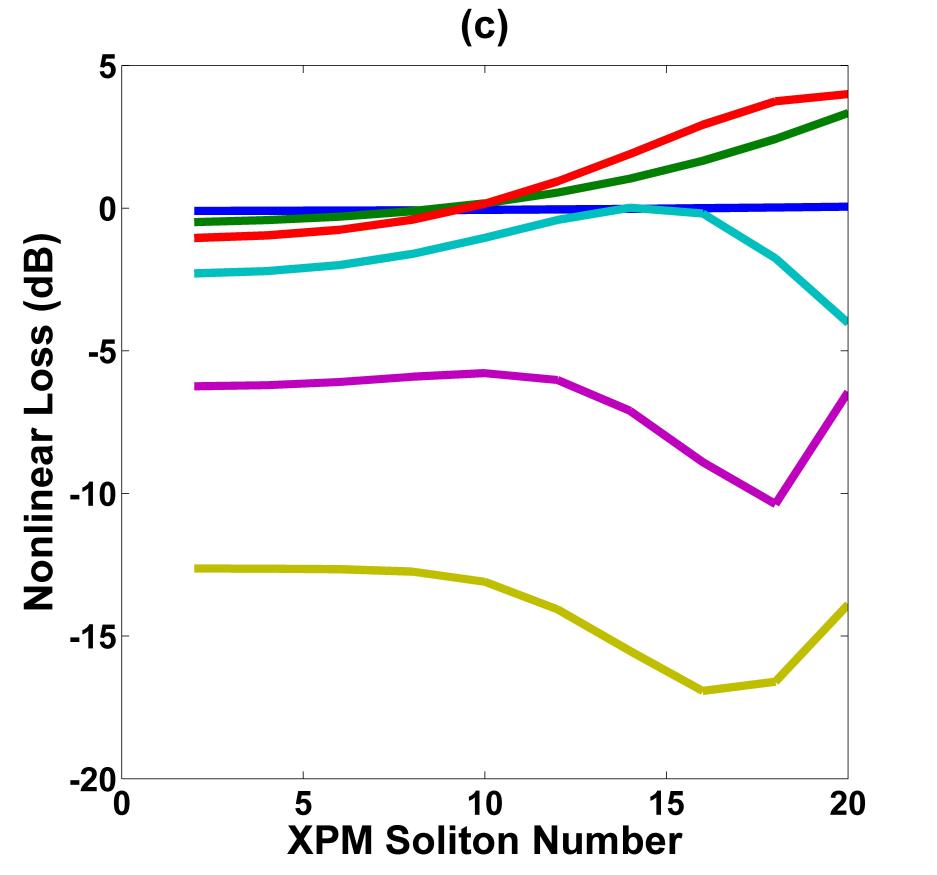}
\end{minipage}
\caption{Results of numerical XPM studies, with a low-energy C-band pulse coupled with a high-energy 2300 nm \emph{pump} pulse: (a) the signal pulse compression factor, (b) the output signal pulse TBP, and (c) the nonlinear peak intensity loss (dB).}
\label{fig:fig3}
\end{figure} 

\section{Conclusion}
The simulations conducted have demonstrated the physical ability of soliton pulse compression and propagation via XPM in a silicon PhCWG with TPA and FCA present at both the pump and signal wavelengths.  Use of a pump pulse and XPM to focus the pulse enables the soliton phenomena to occur on pulses with much lower intensities and thus much lower TPA.  One challenge that was successfully overcome was to develop a NLSE Split-Step Fourier algorithm that would take into account the differences in group velocity and pulse walk-off that will occur between two pulses of different wavelengths.  The simulation demonstrated that through the use of XPM, temporal soliton pulse propagation can be realized over practical length scales with low-intensity ultrashort pulses in a silicon PhCWG with TPA.  This phenomenon might be practically used to sustain a data pulse within a silicon waveguide, a capability with numerous potential technological applications.  \\

Sources of funding for this effort include Navy Air Systems Command (NAVAIR)-4.0T Chief Technology Officer Organization as an Independent Laboratory In-House Research (ILIR) Basic Research Project (Nonlinear Analysis of Ultrafast Pulses with Modeling and Simulation and Experimentation); the National Science Foundation (NSF), the National Science Foundation of China (NSFC) award (61070040 and 60907003), and the Science Mathematics And Research for Transformation (SMART) fellowship.  The author's thank Chee Wei Wong, Jiali Liao, James McMillan, Tingyi Gu, Kishore Padmaraju, Noam Ofir, and the laboratory of Karen Bergman for fruitful discussions.


\begin{thebibliography}{99}

\bibitem{1} “Nonlinear Fiber Optics,” 4$\mathrm{^{th}}$ Ed. Govind Agrawal.

\bibitem{2} "Fundamentals of Photonics.” 2$\mathrm{^{nd}}$ Ed.  Saleh, Teich.  

\bibitem{3} "Numerical Techniques in Electromagnetics, 2$\mathrm{^{nd}}$ Ed." Matthew Sadiku. CRC Press, 2001. 

\bibitem{4} "Engineering Optics with Matlab." Ting-Chung Poon, Taegeun Kim. World Scientific, 2006. 

\bibitem{6} “Photonic Crystals, Molding the Flow of Light, 2$\mathrm{^{nd}}$ Ed.” Joannapoulos, Johnson, Winn, Meade. Princeton U Press 2008.

\bibitem{8} Panoiu, McMillan, Wong. “Influence of the group-velocity on the pulse propagation in 1D silicon photonic crystal waveguides.”  Applied Physics A, Materials Science and Processing.  Vol. 16, No. 1, January/February 2010.  

\bibitem{7} Colman, Husko, Combrie, Sagnes, Wong, De Rossi.  “Temporal solitons and pulse compression in photonic crystal waveguides,” Nature Photonics, 4, 862 (2010). 

\bibitem{9} C. A. Husko, A. De Rossi, and C. W. Wong, “Effect of multi-photon absorption and free carriers on self-phase modulation in slow-light photonic crystals.” Optics Letters, Vol. 36, No. 12. June 15, 2011

\bibitem{21} R.S. Tucker, "The role of optics in computing", Nature Photonics, no.4, p. 405.

\bibitem{22} Steve Paley.  "Now Just a Blinkin' Picosecond!" {\url{http://science.nasa.gov/science-news/science-at-nasa/2000/ast28apr_1m/}}

\bibitem{5} J. F. McMillan, M.Yu, D.-L. Kwong, and C. W. Wong.  “Observations of four-wave mixing in slow-light silicon photonic crystal waveguides,” Optics Express 18, 15484 (2010)

\bibitem{13} Q Lin, O Painter, G Agrawal.  “Nonlinear optical phenomena in silicon waveguides: Modeling and applications.”  Optics Express, Volume 15 No 25, 10 December 2007.  

\bibitem{14} Dawn Tan, Pang Sun, Yeshaiahu Fainman.  “Monolithic nonlinear pulse compressor on a silicon chip.”  Nature Communications.  16 November 2011.  

\bibitem{15} Ding, Gorbach, Wadswarth, Knight, Skryabin, Strain, Sorel, and De La Rue.  “Time and frequency domain measurements of solitons in subwavelength silicon waveguides using a cross-correlation technique.”  Optics Express, 2010. 18(25).

\bibitem{19} M. Notomi, A. Shinya, S. Mitsugi, E. Kuramochi, and H. Ryu, "Waveguides, resonators and their coupled elements in photonic crystal slabs," Optics Express 12(8), 1551-1561 (2004). 

\bibitem{20} T. Gu, N. Petrone, J.F. McMillan, A. Van der Zande, M. Yu, G. Q. Lo, D. L. Kwong, J. Hone, and C. W. Wong, "Regenerative oscillation and four-wave mixing in graphene optoelectronics," Nature Photonics 6, 554 (2012). 

\bibitem{28} M. Dinu, F. Quochi, and H. Garcia.  "Third-order nonlinearities in silicon at telecom wavelengths."  Applied Physics Letters, 18(25), 5 May 2003. 

\bibitem{26} Nagle, R. Kent, Edward B. Saff, and Arthur David Snider, ${\emph{Fundamentals of Differential Equations and Boundary Value Problems}}$, 5$\mathrm{^{th}}$ Ed., Addison Wesley.

\bibitem{27} Richard Haberman. ${\emph{Applied partial differential equations, with Fourier series and boundary value problems}}$. Pearson Prentice Hall, Upper Saddle River, N.J.

\bibitem{10} "Introduction to Solid State Physics, 8$\mathrm{^{th}}$ Ed." Kittel, Charles. Wiley, 2004.

\bibitem{11} Hernando Garcia and Ramki Kalyanaraman.  “Phonon-assisted two-photon absorption in the presence of a dc-ﬁeld: the nonlinear Franz–Keldysh effect in indirect gap semiconductors.”  Journal of Physics B: Atomic, Molecular and Optical Physics. 

\bibitem{12} X. Yang, C. W. Wong. “Coupled-mode theory for stimulated Raman scattering in high-Q/Vm photonic band gap defect cavity lasers.”  Optics Express 15, 4763 (2007)

\bibitem{25} Ultrashort Free-Carrier Lifetimes in Low-Loss Silicon Nanowaveguides. Amy C Turner-Foster,Mark Foster, Cornell University. 15 February 2010 / Volume 18, Number 4, Optical Express 3582.

\bibitem{23} RP Photonics Encyclopedia of Laser Physics and Technology, "Two-photon Absorption."  {\url{http://www.rp-photonics.com/two_photon_absorption.html}}

\bibitem{16} Wei Hseih, Xiaogang Chen, Jerry Dadap, Nicolae Panoiu, and Richard Osgood.  "Cross-phase modulation-induced spectral and temporal effects on co-propagating femtosecond pulses in silicon photonic wires."  Optics Express, 2007.  15(5).  

\bibitem{29} Slusher, Eggleton.  \emph{Nonlinear Photonic Crystals}.  (Springer 2003). 

\bibitem{17} Synchronously Pumped OPO, Angewandte Physik and Elektronik (APE) GmbH, PP Automatic.  OPO PP Auto Manual \#122970

\bibitem{18} R\"{u}diger Paschotta.  "Optical Parametric Oscillators." {\url{http://www.rp-photonics.com/optical_parametric_oscillators.html}}

\bibitem{24} Operator's Manual, The Coherent Mira Optima 900-P Laser

\end{thebibliography}
\end{document}